\documentclass[10pt, conference, compsocconf]{IEEEtran}
\usepackage{color}
\usepackage{multicol}
\usepackage{graphicx}
\usepackage{multirow}
\usepackage{booktabs}
\usepackage[table]{xcolor}

\begin{document}

\title{A Comprehensive Survey of MAC Protocols\\ for Wireless Body Area Networks}

\author{\IEEEauthorblockN{A. Rahim, N. Javaid, M. Aslam, Z. Rahman, U. Qasim$^{\ddag}$, Z. A. Khan$^{\S}$\\}

        $^{\ddag}$University of Alberta, Alberta, Canada\\
        Department of Electrical Engineering, COMSATS\\ Institute of
        Information Technology, Islamabad, Pakistan. \\
        $^{\S}$Faculty of Engineering, Dalhousie University, Halifax, Canada.
        }

\maketitle

\begin{abstract}
In this paper, we present a comprehensive study of Medium Access Control (MAC) protocols developed for Wireless Body Area Networks (WBANs). In WBANs, small battery-operated on-body or implanted biomedical sensor nodes are used to monitor physiological signs such as temperature, blood pressure, ElectroCardioGram (ECG), ElectroEncephaloGraphy (EEG) etc. We discuss design requirements for WBANs with major sources of energy dissipation. Then, we further investigate the existing designed protocols for WBANs with focus on their strengths and weaknesses. Paper ends up with concluding remarks and open research issues for future work.\\


\end{abstract}

\IEEEpeerreviewmaketitle

\section{Introduction}
Number of small and smart devices increases due to advancement in wireless and storage technologies. These small devices are capable of long time health monitoring with in hospital or outside. Wireless Body Area Networks (WBANs) enable us to use portable, small and lightweight sensor nodes for long time health monitoring. Using sensing capabilities, these small energy constrained devices measure human body parameters and communicate with some external monitoring station for diagnose or prescription from a physician. Data streaming from human body to monitoring station using wireless communication channel is an energy consuming process. Low power signal processing and energy efficient communication mechanisms prolong lifespan of these small devices. For Low-Rate Wireless Personal Area Networks (LR-WPANs), IEEE 802.15.4 defines specification for Physical Layer and Data Link Layer [1].

In WBANs, sensor nodes of small size with low power and limited computational capabilities are attached or implanted to human body for measurement of  physiological signs. These physiological signs include; respiratory patterns, heartbeat, temperature, posture, breathing rate, ElectroCardioGram (ECG), ElectroEncephaloGraphy (EEG) and many more. Transmission data rates for these physiological parameters vary from 1Kbps to 1Mbps. Sensor nodes collect information from human body and communicate with a central device called Coordinator.

Energy efficiency is the most important requirement of a good MAC protocol for WBANs. To improve energy efficiency of WBANs, a versatile MAC protocol should have the capabilities to reduce power dissipation due to collision of packets, overhearing of nodes, idle listening to receive probable data packets  and  control packet overhead of communication. Similarly Quality of Service (QoS) is an important goal to achieve in WBANs. This includes latency, jitter, guaranteed communication and security.

For fair access of shared medium, MAC protocols for Wireless Sensor Networks (WSNs) and other short range wireless technologies use Time Division Multiple Access (TDMA) or Carrier Sense Multiple Access with Collision Avoidance (CSMA/CA). Due to complex hardware and high computational power requirements, Frequency Division Multiple Access (FDMA) and Code Division Multiple Access (CDMA) are not suitable approaches for medium access in sensor networks [5]. CSMA/CA approach out performs in dynamic networks. It is presumed that WBANs are not dynamic. TDMA approach is well suited for WBANs. However, TDMA-based MAC protocols require extra energy consumption for synchronization. Comparison of CSMA/CA and TDMA is given in Table I.

\begin{table}[htbp]
  \centering
  \caption{Comparison of CSMA/CA and TDMA }
    \begin{tabular}{lll}
    \toprule
    Feature & CSMA/CA & TDMA \\
    \midrule
    Power Consumption & High  & Low \\
    Bandwidth utilization & Low   & Maximum \\
    Traffic level support & Low   & High \\
    Mobility(Dynamic) & Good  & Poor \\
    Synchronization & N/A   & Necessary \\
    \bottomrule
    \end{tabular}%
  \label{tab:addlabel}%
\end{table}%

%


We organize our work in this paper as follows. In subsequent section, we discuss design requirements for WBANs, however, section III presents major sources of energy dissipation. In section IV, we present existing MAC protocols with their pros and cons. Section V provides a brief discussion with a number of open research issues for design of efficient and  reliable MAC protocols for WBANs. Finally, section VI concludes the research work carried out in this paper.

\section{Design Requirements for WBANs}

In WBANs, sensor nodes collect critical and non-critical information from different parts of patient body and communicate with coordinator. Latency and transmission reliability are important requirements for effective patient health monitoring systems. Similarly for long time monitoring, WBANs required high energy efficiency and scalability at Level 1.

\subsection{Energy Efficiency}

Energy efficiency is the first goal to achieve in WBANs since sensor nodes are small and battery operated. For long time patient monitoring, it is an obligatory goal to play down energy dissipation at Level 1 as much as possible. Multiple and dynamic power management schemes can be used to prolong lifespan of sensor nodes. In WBANs, sensor node's transceiver is one of the dominant energy dissipation source. Optimization of PHYsical (PHY) and MAC layer processes result in reduced power consumption of transceiver. PHY layer layer has some limitation for power optimization. However, MAC layer provides higher level of energy savings by introducing multiple transmission scheduling schemes, optimal packet structure, smart signaling techniques and enhanced channel access techniques.

\subsection{Reliability}

Reliability of WBANs depends upon transmission delay of packets and packet loss probability. Packet transmission procedures at MAC layer and Bit Error Rate (BER) of channel influence packet loss probability. Appropriate channel access techniques, packet re-transmission schemes, packet size, and enhanced scheduling schemes at MAC layer improve reliability.

\subsection{Scalability}

Scalability is the essential requirement for WBANs. Number of nodes, to collect life critical and non-critical information, varies according to patient monitoring requirements.  Easily configuration of WBANs by adding or removing sensor nodes is required to support scalability. MAC layer has potential to achieve scalability.

\subsection{Quality of Service (QoS)}

MAC layer play a vital role to achieve high QoS. Medium access techniques at MAC layer like TDMA and polling put forward deterministic packet loss, packet delay. However, contention based protocols like CSMA allocates transmission channel to node only when it is free and node has data to transmit. Random access techniques result in variable packet loss and packet delay. Adaptive sleep cycles in contention based protocols enhance energy efficiency at the cost of increase latency and packet drops.

\section{Sources of Energy Dissipation in WBANs}

Sensor nodes have small batteries with limited power capabilities. Replacement or recharging of batteries by energy scavenging is not possible. Due to limited energy resources, power consumption of sensor nodes needs to be controlled tightly. Thus minimization of energy consumption is a major issue in WBANs.  Power consumption of sensor nodes can be decreased with  low power MAC protocols.  Collision of packets, overhearing of nodes, idle listening to receive possible data packets , communication control packet overhead, packet forwarding and transceiver state switching are the foremost sources of energy dissipation in Wireless Sensor Networks (WSNs). In [12] authors identify the first four sources of energy dissipation.

Transmission of data packets on single channel by two or more sensor nodes simultaneously results in packet collision. Collision of packets occurs at receiver end. These packets are dropped and sender nodes retransmit these packets. Re-transmission of dropped packets results in extra energy dissipation. In overhearing, sensor nodes receive packets that are destined for other nodes. Those received packets are dropped and energy is dissipated. In idle listening, nodes listen to idle channel to receive possible packets transmitted by other nodes which results in extra energy consumption. If control packets used in communication are maximum effective throughput decreases. Transmission and reception of these maximum control packets consume more energy. Energy is consumed in packet forwarding, when router nodes consume energy to forward a data packet from source to destination. However, energy consumption due to packet forwarding is ignored in WBANs due to single-hop communication in star topology. The last source is state switching, which occurs when a sensor node switch its transceiver from sleep mode to active mode for data transmission and then back to sl eep mode to avoid idle listening and overhearing. Frequent switching of transceiver is also energy consuming. Energy efficiency can be improved by avoiding such energy  wastage sources in efficient way.

\section{MAC Protocols for WBANs}

In this section, we discuss some of well known existing MAC protocols proposed for WBANs. This discussion covers the pros and cons of these proposed protocols in context of energy minimization and resource utilization. The following subsections provide detail operation of these protocols with emphasize on energy consumption. We also discuss, how these protocols tackle energy inefficiency sources like collision, idle listening, overhearing and control packet overhead which are widely addressed in literature.

\subsection{IEEE 802.15.4 MAC Protocol}

IEEE 802.15.4 is designed for low data rate wireless applications [1]. This protocol operates in three frequency bands: 868 MHz, 915 MHz and 2.4 GHz frequency bands. These frequency bands are further divided into 27 sub-channels i.e., 2.4 GHz frequency band is divided into 16 sub-channels, 915 MHs frequency band into 9 sub-channels and one sub channel in 868 MHz frequency band. Two operational modes are defined for IEEE 802.15.4: beacon enabled mode and non-beacon enabled mode.

In beacon enabled mode, coordinator controls device synchronization, association and data transmission using periodic beacons. Beacon enabled mode use a super frame. This super frame consists of active and inactive periods. Active period of superframe is divided into three parts: Contention Access Period (CAP) using slotted CSMA/CA, beacon and a Contention Free Period (CFP).  A maximum of seven Guaranteed Time Slots (GTS) are assigned to end nodes to accommodate time critical data in CFP. This mode of operation of IEEE 802.15.4 is not suitable for WBANs due to its asymmetric transmission support.
Non-beacon enabled mode of IEEE 802.15.4 uses un-slotted CSMA/CA. Authors in [2], analyze slotted and un-slotted CSMA/CA and presented their results. These results show that un-slotted mechanism out performs well in terms of bandwidth utilization and latency as shown is Fig. 1. However, in non-beacon enabled mode, Clear Channel Assessment (CCA) leads to high energy consumption.

\begin{figure}[ht]
\begin{center}
\vspace{-0.4cm}
\includegraphics[scale=0.17]{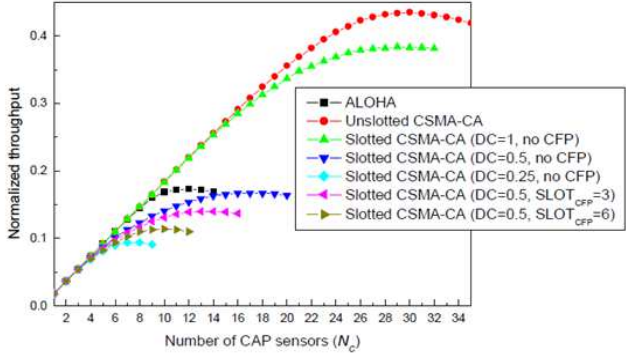}
\vspace{-0.4cm}
\caption{Normalized Throughput Versus NC}
\end{center}
\vspace{-0.6cm}
\end{figure}

\subsection{Battery-aware TDMA Protocol}

In [3], authors propose a battery-aware TDMA based MAC protocol with cross-layer design to maximize network life. This protocol takes the following parameters into account for medium access: electrochemical properties of battery, time-varying wireless fading channel, and packet queuing characteristics.  Operation of this protocol is similar to IEEE 802.15.4 beacon enabled mode, where nodes listen periodically to beacons from coordinator. The time axis is divided into three parts; beacon slot, active time slots and inactive period as shown in Fig. 2 [3]. The frame length is adaptive and can be changed according to application requirements. Sensor nodes  wake up at beginning of beacon period. Each node has its own distinct time slot $T_s$ to transmit data in active period after receiving beacon. To avoid extra energy consumption, nodes remain in sleep mode for inactive time. This protocol prolongs lifespan of wireless sensor nodes. Reliable and timely delivery of packets is achieved using GTS.  However, there is no mechanism defined for emergency data. Similarly holding of packets in buffer for long time, leads to high average delay and packet drop rate.

\begin{figure}[ht]
\begin{center}
\vspace{-0.4cm}
\includegraphics[scale=0.2]{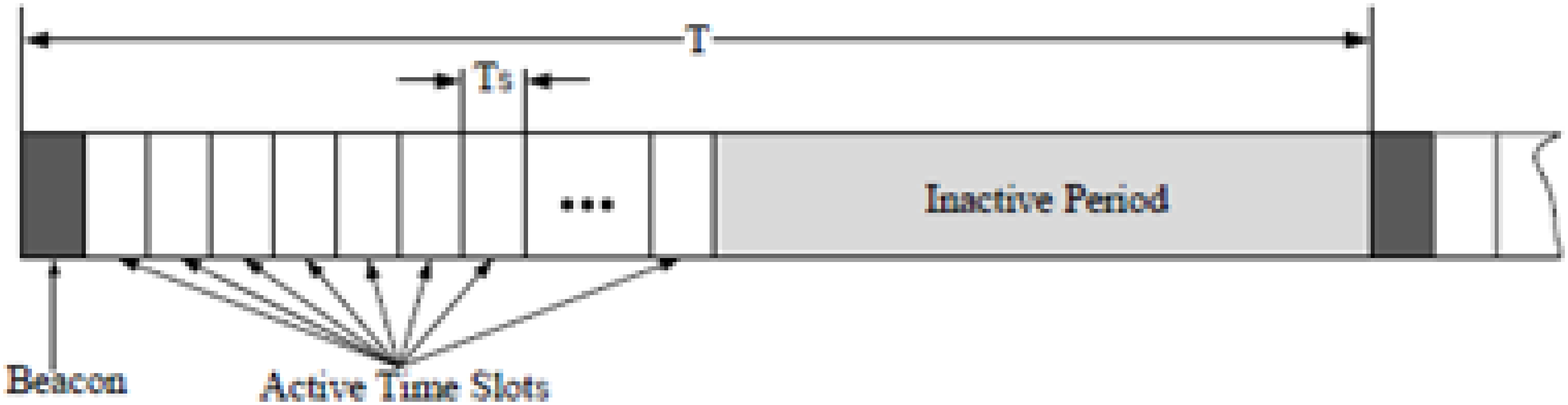}
\vspace{-0.5cm}
\caption{TDMA Frame Structure}
\end{center}
\vspace{-0.5cm}
\end{figure}

\subsection{Priority Guaranteed MAC Protocol}

In [4], authors propose a priority-guaranteed MAC protocol.  This protocol uses a new superframe structure as shown in Fig. 3. The active period is divided into five parts; a beacon, Control Channel AC1, Control Channel AC2, Time Slot Reserved for Periodic (TSRP) traffic, and Time Slot Reserved for Bursty (TSRB) traffic.  AC1 is used for uplink control of life-critical medical application while AC2 is used for uplink control of Consumer Electronics (CE) applications. Randomized ALOHA is used for these two control channels. Proposed protocol is based upon TDMA approach to assign Guaranteed Time Slots (GTS) within two data channels TSRP and TSRB. These time slots are allocated on-demand to using control channels. As shown in simulation results [4], this protocol out performs than IEEE 802.15.4 in terms of energy consumption. However, complex superframe structure and inadaptability to emergency traffic are major drawbacks of this protocol.

\begin{figure}[ht]
\begin{center}
\vspace{-0.4cm}
\includegraphics[scale=0.18]{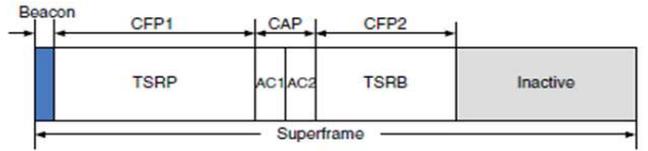}
\vspace{-0.5cm}
\caption{Superframe Structure of Priority-Guaranteed MAC }
\end{center}
\vspace{-0.6cm}
\end{figure}

\subsection{Energy-Efficient Low Duty Cycle MAC Protocol}

Authors propose a new MAC protocol based upon static nature of BAN [5]. TDMA approach is used for streaming large amount of data. Static nature and TDMA approach are being utilized efficiently to maximize  network life. In target topology a Master Node (MN) collects data from on body nodes and communicates with a Monitoring Station (MS). Received data is being analyzed by MS while the on-body network coordination and synchronization is being performed by MN.  As shown in Fig. 4, total frame is divided into multiple time slots. Time slots S1 to Sn are allocated to sensor nodes while time slots RS1 to RS2 are reserved which are being assigned when requested. Number of these extra time slots depends upon targeted packet drop, packet error rate and number of sensor nodes.

\begin{figure}[ht]
\begin{center}
\vspace{-0.4cm}
\includegraphics[scale=0.18]{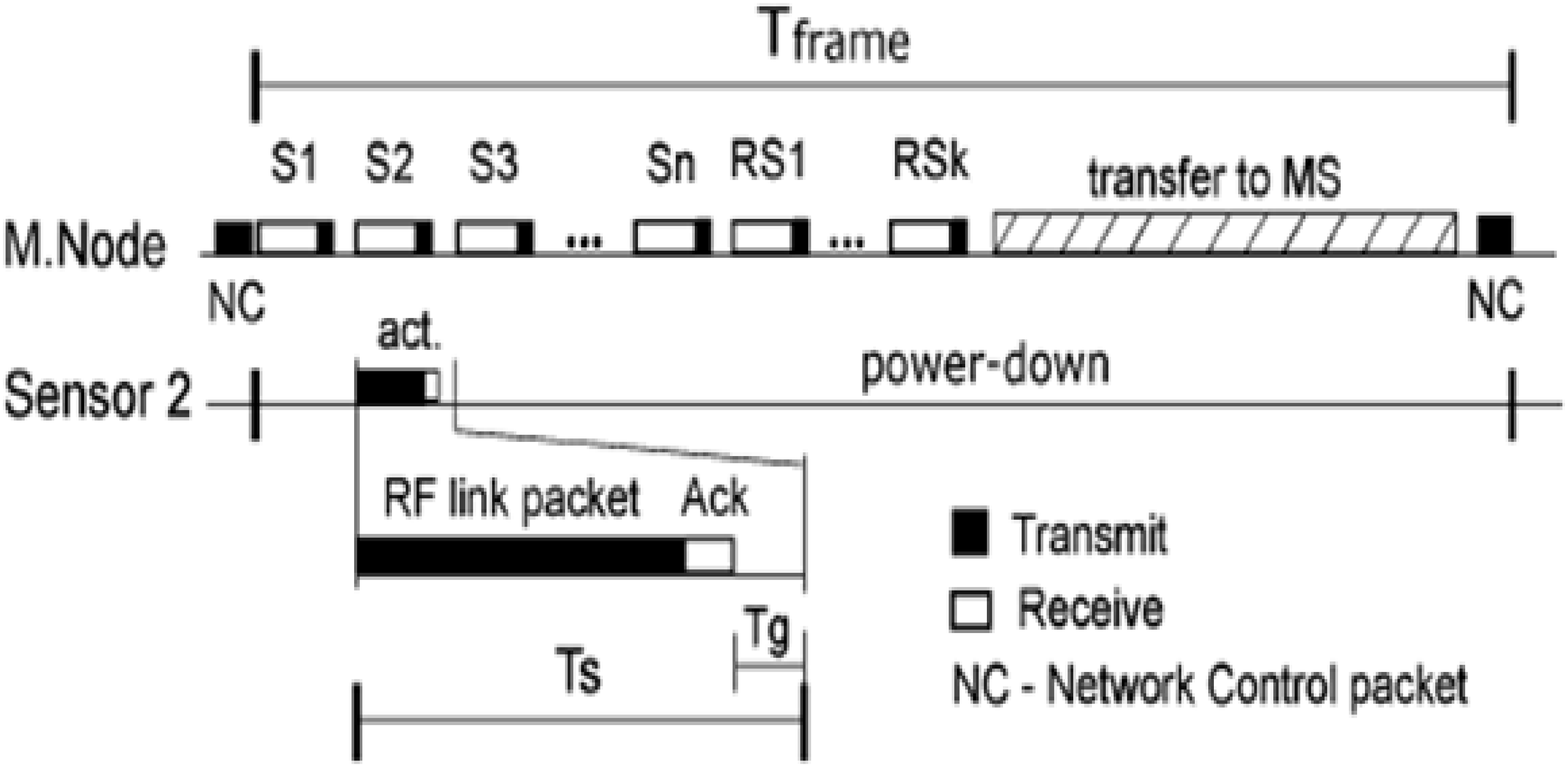}
\vspace{-0.4cm}
\caption{TDMA Frame Structure}
\end{center}
\vspace{-0.6cm}
\end{figure}

To avoid collision/overlapping of packets transmission due to clock drifts, guard band time is inserted between two consecutive time slots. Two types of communication models are being discussed. First, MN has one transceiver. In this case, enough time is reserved for communication of MN with MS. In second case, where the MN has two transceivers, simultaneously communication of MN with MS and sensor nodes is possible. The communication uses different physical layer communication models for transparency.

From energy consumption analysis in [5], proposed protocol out performs in term of energy for high communication data rates as well as for short burst of data. However, this protocol uses a Network Control (NC) packet for periodic synchronization after $N$ number of time frames which leads to an extra consumption of energy. Other shortcoming includes; fixed frame structure based on pure TDMA, no CAP to accommodate small burst of data,  and no mechanism is defined for on-demand traffic.

\subsection{A Power-efficient MAC Protocol for WBANs}
In [6], authors propose a new mechanism at MAC layer to accommodate normal, emergency, and on-demand traffic. For reliable transmission two wakeup mechanisms are defined: traffic-based wakeup mechanism for transmission of normal traffic and wakeup radio mechanism for emergency/on-demand data transmission. Normal traffic is generated periodically by sensor nodes to monitor routine physiological parameters. Unpredictable emergency traffic is initiated by on-body sensor nodes in emergency situation. However, coordinator generates on-demand traffic to acquire information from sensor nodes. A new superframe structure is defined where, time axis is divided into three parts: a beacon message, a Configurable Contention Access Period (CCAP) to accommodate short burst of data, and a Contention Free Period (CFP) where Guaranteed Time Slots (GTS) are assigned to end nodes for collision free communication. In CCAP, proposed protocol uses slotted ALOHA. Superframe structure for this protocol is shown in Fig. 5 [6].
\begin{figure}[ht]
\begin{center}
\vspace{-0.4cm}
\includegraphics[scale=0.16]{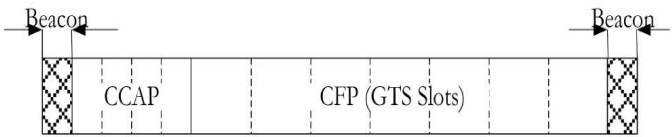}
\vspace{-0.8cm}
\caption{Superfame Structure }
\end{center}
\vspace{-0.6cm}
\end{figure}

Coordinator organizes traffic-based wakeup table according to application. Periodic sleep/wakeup mode avoids unnecessary energy dissipation due to idle listening and overhearing. To compensate clocks drift at coordinator and sensor nodes, sensor nodes wake up in advance for a time period of $T_K =2 \theta T_W $ where $T_W$ is the beacon period. For emergency traffic sensor nodes send wake up radio signal to coordinator while coordinator sends a wake up signal to sensor nodes for on demand traffic.  Simulation results based upon Monte Carlo method for poisson and deterministic traffic. Performance of proposed protocol in terms of energy and delay are compared with that of WiseMAC [7], where it performs better. However, use of Low Power Listening (LPL) is not an optimal choice for implanted and on-body sensor nodes communication due to strict power capabilities.

\subsection{Energy Efficient Medium Access Protocol}

In [8], authors propose a new MAC protocol based upon centrally controlled wakeup and sleep mechanisms to maximize energy efficiency.  Some upper layer functionalities are incorporated to reduce power dissipation due to overhead. This protocol is based upon basic assumption of sensor nodes with a star topology where a central node (master node) coordinates with on-body/implanted sensor nodes (Slave nodes). Maximum number of slave nodes for a single master node is defined to be 8. Due to high power and computational capabilities, more activities and processes are assigned to central node.

Basic operation of this MAC protocol involves three processes. First one is link establishment, where a slave node wants to join a cluster. After successful link establishment, each node is assigned with a unique sleep time to avoid idle listening and overhearing. Second one is the wakeup service process, where master and slave nodes communicate. Exception process, also called an Alarm process is initiated by slave node to communicate with master node for emergency data. For guaranteed and reliable communication, a novel concept of Wakeup Fallback Time (WFT) is introduced. In case of failure in assigned wakeup process, sensor node enters into sleep mode for a specific time interval calculated by WFT. During this sleep time, sensors node buffers data packets for future communication. Similarly, master node also goes into sleep mode set by WFT if it fails to communicate with slave nodes. Overlapping of time slots is being avoided by this mechanism.

From simulation results for different applications such as glucose monitoring, human body temperature, EEG, and ECG, power consumption depends on sleep interval and number of retransmissions as shown in Fig. 6. The centrally controlled process reduces efficiently the extra energy consumption due to idle listening and overhearing. However, implementation of this protocol is highly complex and has no proper mechanism to handle on-demand traffic. Other drawbacks include: limitation of nodes in one cluster, communication is only initiated by mater node and only one node goes into link establishment process at a time.

\begin{figure}[ht]
\begin{center}
\vspace{-0.4cm}
\includegraphics[scale=0.17]{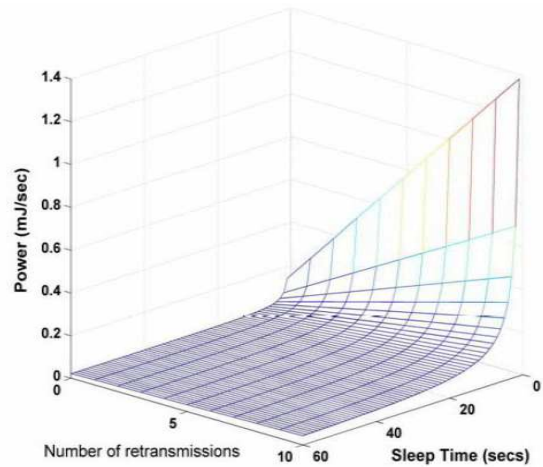}
\caption{Power Compared to Sleep Time and Number of Retransmits}
\vspace{-0.4cm}
\end{center}
\vspace{-0.4cm}
\end{figure}

\subsection{BodyMAC}
In [9], authors propose a TDMA-based MAC protocol where they define uplink and downlink subframes to facilitate sleep mode with emphasize on energy minimization. Nodes remain in sleep mode when they have no data to send. Sleep mode performs well for low duty cycle sensor nodes. Different data communication models are accommodated using 3 bandwidth management procedures; \emph{Burst Bandwidth procedure, Periodic Bandwidth procedure and Adjust Bandwidth procedure}. This efficient and flexible bandwidth management procedure improves network stability and improves transmission of control packets.

\begin{figure}[ht]
\begin{center}
\vspace{-0.4cm}
\includegraphics[scale=0.25]{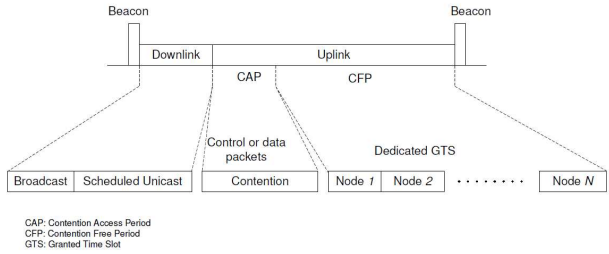}
\vspace{-0.4cm}
\caption{BodyMAC Frame Structure}
\end{center}
\vspace{-0.5cm}
\end{figure}

As shown in Fig. 7 [9], MAC frame is divided into three parts; a beacon, a downlink and uplink. Synchronization is archived by beacon. To accommodate on demand traffic, downlink is used for data communication from coordinator node to sensor nodes. However, uplink frame is divided into Contention Access Period (CAP) and Contention Free Period (CFP). CAP is based on CSMA/CA, where nodes compete to send control packets to coordinator for Guaranteed Time Slots (GTS). However, nodes can also communicate for small data packets during CAP. Coordinator assigns GTS to sensor nodes in CFP to avoid collision. Communication using CFP improves energy effecting. However, for uplink frame in CAP, CSMA/CA ends up with high energy consumption due to Clear Channel Assessment (CCA) and collision issues.

\subsection{MedMAC}
In [10], authors propose Medical Medium Access Control (MedMAC) protocol for WBANs to improve channel access mechanism and reduce energy dissipation. MedMAC using TDMA approach for time slots assignment to end nodes for data communication. However, these assigned time slots are of variable length and vary according to sensor nodes requirements. A novel scheme is introduced for synchronization. MedMAC uses multi-superframe structure, where beacons are used for synchronization as shown in Fig. 8 [10].  For network initialization, emergency traffic, and low data communication it uses an optimal contention period.

\begin{figure}[ht]
\begin{center}
\includegraphics[scale=0.25]{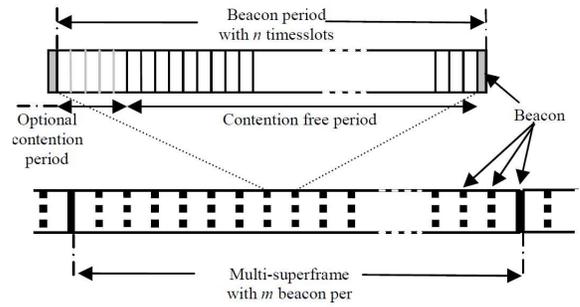}
\caption{Multi-Superframe Structure for MedMAC Protocol}
\end{center}
\vspace{-0.6cm}
\end{figure}

To maintain clock synchronization of nodes and coordinator, MedMAC uses timestamp scavenging with Adaptive Guard Band Algorithm (AGBA). Collision of data packets is avoided using unique GTS for each sensor node. Similarly AGBA maintain synchronization of devices to avoid collision due to clocks drift. Using AGBA guard band time is inserted between two consecutive time slots. This guard band time is adjustable and based on clock drift of devices. Drift Adjustment Factor (DAF) monitors guard band and avoid waste of bandwidth using extra guard bands.

Authors use OPNET for simulation. They compare performance of MedMAC with that of IEEE 802.15.4 with respect to power dissipation. For applications with low data rates like  pulse (8 bps),respiration (640 bps), and temperature (16 bps), and medium data like ECG, simulation are performed. From simulation results in [10], it is concluded that it out performs than IEEE 802.15.4 with respect to energy consumption. Collision is being avoided using GTS. However, MedMAC takes low data traffic into consideration which is not suited in WBANs where date rates for wearable and implanted sensors may be high.

\subsection{Heartbeat-Driven MAC Protocol}
In [11], authors propose a TDMA based protocol for WBANs with utilization of Heartbeat-Rhythm for synchronization. Network topology for proposed protocol is star topology where a central node coordinates network. To avoid collision, H-MAC assigns dedicated time slots to sensor nodes for communication. Using Heartbeat Rhythm, H-MAC maintains synchronization required for TDMA approach without using periodic control messages. This mechanism leads to minimize overall energy consumption. Each biosensor extracts Heartbeat Rhythm information from its sensory data. For detection of peaks in the heartbeat rhythm, authors use the algorithms proposed in [12,13]. Synchronization is archived by these peaks. H-MAC uses the peek intervals for data communication. Time slots assignment and frame cycles for synchronization are calculated by coordinator. Coordinator also utilizes Heartbeat Rhythm information from its own sensory data.

From simulation results, H-MAC prolongs networks life as compared to Lightweight MAC (L-MAC) [14] and Sensor MAC (S-MAC) [15]. This efficiency is achieved by TDMA approach, where collisions are avoided by dedicated time slots and reduced idle listening. Replacement of traditional synchronization with Heartbeat Rhythm pattern also reduces energy consumption. However, Heartbeat Rhythm is not accessible by all sensors like accelerometer. In such cases devices can not by synchronized. Integration of accelerometer with other sensors or facilitating accelerometer to access heartbeat leads to complexity. Similarly insertion of guard band time to avoid collisions leads to minimize bandwidth utilization.

\section{Discussion and Open Research Issues}

Energy efficiency is one of the main goals to achieve in WBANs for mobile and ubiquitous health monitoring with critical and non-critical conditions. Current research work for energy minimization is focused at MAC layer. However, other areas such as network layer and cross layer design need to be consider for energy minimization. In cross layer design we can improve energy efficiency by integrating two or more protocol layers from communication protocol stack. Therefore, research work using cross layer approach will be prominent field to minimize energy consumption. Similarly Radio Frequency (RF) communication, antenna design, and propagation modules effect performance of WBANs. Other issues for researcher to be consider includes mobility of on, in or around human body sensor nodes, transparency at MAC layer, interoperability, security and QoS.

CDMA, FDMA, CSMA, and TDMA are multiple approaches for medium access. However, each of them has some advantages and disadvantages. Collision free communication is achieved by CDMA, but high computational and power requirements are major obstacles for implementation in WBANs where sensor nodes have limited computational capabilities with constrained power. Hardware complexity required for FDMA, to achieve collision free channel access, makes FDMA an inappropriate solution for WBANs. CSMA based MAC protocols provide promising results such as low delay, reliable communication, and  simple implementation procedure in small dynamic networks. However, additional energy consumption for collision detection or collision avoidance, and protocol overhead are major shortcomings of CSMA. TDMA-based MAC protocols are contention free; nodes transmit data in predefined time slots to avoid packet collision. For small networks with low mobility and small number of sensor nodes and periodic data generation, TDMA is the best approach for medium access. However, strict synchronization requirement, non-adaptability and scalability are some issues faced by TDMA. Based on topology and limited number of nodes in WBANs, TDMA could be considered most suitable solution for medium access in WBANs.

Energy efficiency is of utmost importance in WBANs. For high energy efficiency, a number of protocols has been proposed. However, MAC protocols specifically for WBANs need to be developed. Aim of these protocols would be to avoid energy dissipation due to collision, overhearing and idle listening with reduced control packet overhead and implementation complexities. Fairness at MAC layer, high bandwidth utilization, reliable communication, minimum delay, and reduced synchronization cost are other objectives for multipurpose efficient MAC protocol. Proposed protocol should have capabilities to accommodate communication of normal, emergency, and on-demand traffic. However, selection of MAC protocols is application and hardware dependent. This may be one of the reasons that no proposed protocol is accepted as a standard for WBANs so far.

\section{Conclusion}

Aim of this research work is to analyze existing MAC protocols for WBANs with emphasis on energy minimization. These protocols are being developed to prolong lifespan of WBANs, reliable communication, flexibility, fair management,  and QoS. However, MAC protocols based on random access and LPL are unable to accommodate emergency and on-demand traffic. On the other hand, TDMA is a vital approach for medium access to be used in WBANs.
Majority of existing MAC protocols based on TDMA approach. Each of them  has some advantages and disadvantages discussed above. Due to diverse application requirement and hardware constrains, no one protocol is being accepted as a standard. A new protocol needs to be developed to achieve requirements of WBANs like energy efficiency, scalability, fairness, reduced implementation complexity, support for divers application, interoperability, reduced synchronization overhead, and QoS.


\end{document}